\documentstyle[12pt]{article}
\title{Exotic smoothness and particle 
physics.}
\author{J. S\l adkowski *\\
\sl Institute of Physics, University of Silesia,\\
\sl ul. Uniwersytecka 4, Pl-40007 Katowice, Poland, \\}
\begin{document}
\date{}
\baselineskip24pt
\maketitle
\begin{abstract}
\baselineskip24pt
\ \ \ Short introduction to exotic differential structures 
on manifolds is given. The possible physical context of this mathematical 
curiosity is discussed. The topic is very interesting although 
speculative.
\end{abstract}

\vspace{6mm}

* E-mail: sladk@us.edu.pl
\newpage

\ \ \ Classical differential calculus is defined on a Banach space. It has 
been generalized in two ways: the theory of generalized functions 
(distributions) and the calculus on (differential) manifolds. Both 
generalizations have found profound applications in physics. Here will discuss 
some aspects of the calculus on manifolds (differential geometry) [1-5]. 
Roughly 
speaking, a differential manifold is a topological space $M$ that is locally 
homeomorphic to a Euclidean space (topological vector space in the general 
case). These local homeomorphisms form (provided they fulfil some consistency 
conditions on common domains) what we call an atlas on $M$. A real 
function $f\ :\ M \supset U_{\alpha} \rightarrow {\bf R}$ is said 
to be differentiable 
at $ b 
\in U_{\alpha} $ if its local coordinate 
representation $f_{\alpha}=f\circ
\phi _{\alpha}^{-1}$ is differentiable in the ordinary sense. The union of all 
consistent (that is $\phi _{\alpha} \circ \phi _{\beta}'^{-1}$ is a function of a 
given differentiability class) is called a differential structure on the 
manifold $M$. A function $f\ :\ M \rightarrow M'$ is said to be differentiable 
if its local coordinate representation $\phi _{\beta} \circ f\circ 
\phi _{\alpha}^{-1}$ 
is differentiable. If $f^{-1}$ exists and is differentiable we call it 
difeomorphism and say that $M$ and $M''$ are diffeomorphic (they are identical 
from the differential geometry point of view).  One should ask the fundamental 
question {\it Can two homeomorphic manifolds (that is equivalent as 
topological spaces) support truly different (nondiffeomorphic) differential 
structures?} The answer is yes. What surprising is is the  fact that 
${\bf R^{4}}$, the four-dimensional Euclidean space, 
can be given infinitely many 
nondiffeomorphic (exotic) differential structures! 
In the following we will discuss possible physical consequences of 
this fact. We will start by reviewing some  aspects of exotic differential 
structures on ${\bf R^{4}}$ and other four-dimensional manifolds. 

\ \ \  An exotic ${\bf R_{\Theta}^{4}}$  consists of a set of points 
which can be globally continuously identified with the set four 
coordinates $(x^{1}, x^{2},x^{3},x^{4})$. These coordinates may be 
smooth
locally but they cannot be globally continued as smooth functions 
and no
diffeomorphic image of an exotic ${\bf R_{\Theta}^{4}}$  can be given
such global coordinates in a smooth way. There are uncountable many of
different ${\bf R_{\Theta}^{4}}$. In fact, there is at least a two-parameter 
family of them [5]. C. H.  Brans has proved the 
following theorem [7]:

{\bf Theorem 1.} {\it There exist smooth manifolds which are
homeomorphic but not diffeomorphic to ${\bf R^{4}}$ and for which the
global coordinates $(t,x,y,z)$ are smooth for $x^{2}+y^{2}+z^{2}\geq
a^{2}> 0$, but not globally. Smooth metrics exist for which the
boundary of this region is timelike, so that the exoticness is 
spatially
confined.} \\ 
 
Of course, there are also ${\bf R^{4}_{\Theta}}$  whose 
exoticness
cannot be localized. They might have important cosmological
consequences.  We also have [7]

{\bf Theorem 2.} {\it If $M$ is a smooth connected 4-manifolds and 
$S$ 
is a closed submanifold for which $H^{4}(M,S,{\bf Z})=0$, then any
smooth, time-orientable Lorentz metric defined over $S$ can be 
smoothly
continued to all of $M$.} \\

It can also be proven that if you remove one point from a four-manifold 
then the resulting manifold has its exotic versions [5]. For example, by 
removing a point from ${\bf R^{4}}$ we obtain a manifold that is topologically 
${\bf R\times S^{3}}$ and has exotic differential structures that might  
be very important for cosmologists. 

\ \ \ The discussion of possible physical consequences of the existence of 
exotic differential structures  on some four-manifolds is very difficult 
because we lack such important ``details'' as explicit construction of 
a metric tensor and so on. Nevertheless, some general remarks can be given. 
Brans has even conjectured [7] that a localized exoticness (in the sense 
of the Theorem 1)  
can act as a source for some externally regular field just as matter can. To 
define a ``reasonable'' quantum  field theory on a manifold  we need a notion 
of ``positive frequency'' in the asymptotic past and future (``in'' and 
``out'' states). This is not an easy task for a general spacetime. R. Wald 
has told us [8] how to define a quantum theory in the case of curvature of 
compact support (the spacetime becomes flat in the past and future). This 
means that it might be possible to construct a quantum field theory 
(scattering matrix) on some exotic $R\times S^{3}$. Spacetime of this 
topology arises if we require that all physical fields vanish at spatial 
infinity. Cosmologists  are also exploring such spacetime manifolds. 
    
\ \ \ There is another reason for believing that exotic smoothess 
has a potential physical context. The 28 differential structures on the 
$S^{7}$ and some homeomorphic homogeneous spaces can be distinguished 
by their spectra provided an appropriate metrics is chosen 
(the Pontrjagin forms must vanish). To be more precise we have [9]:   

{\bf Collorary}  {\it 
Suppose $M$ and $M'$ are two topological k-spheres (with codimension one 
metrics), k= 7 or 11. If $M$ and $M'$ are isospectral then they are 
diffeomorphic.} \\ 

Kreck and Stolz have shown [10] that certain Einstein seven-manifolds with $SU(3)
\times SU(2)\times U(1) $ symmetry are distinguished by their spectra [10]. 
They have also given an example of an Einstein manifold with an exotic 
structure admitting again an Einstein metrics. Stolz [11] 
has shown that exotic  
differential structures on some four-manifolds can   
be detected by spectral invariants of the twisted Dirac operator. For example, 
$\eta ({\bf RP^{4}},g,\phi ) \not= \eta ({\bf Q^{4}}, g',\phi ')$ for all 
metrics $g$ and $g'$. Here ${\bf Q^{4}}$ denotes an exotic version of the 
real four-dimensional projective space and $\phi $ the pin-structure. $\eta $ 
is the famous eta invariant (asymmetry of the spectrum 
of the Dirac operator). All 
these examples are important from the Kaluza-Klein or string-inspired models 
because spectra of internal spaces often determine physical data [12, 13]. 

\ \ \ Let us now consider the A. Connes construction of the standard model 
lagrangian [14-17]. The spacetime consists of at least two copies of the 
ordinary spacetime manifold. One may ask if both manifolds carry the same 
differential structure. If not [18] we must impose some consistency 
conditions to make to make the fields defined with respect to different 
differential structures compatible. The simplest and easiest condition 
to fulfil is to demand that field smooth with respect to one differential 
structure must have compact supports (constant function are smooth so 
there is no smoothness problem outside the support).  This means that 
exotic smoothness may be a source of sort of confinement if the Connes 
construction is correct (bag-like structures). This might be important from 
the astrophysical/cosmological point of view (dark matter?). 

\ \ \ Let me conclude this sketchy review by stating 
that exotic smoothness is 
interesting not only as a 
mathematical curiosity but also for its physical context. 
If Nature have not used exotic 
smoothness we should find why. We should also know if and why only one 
of the existing differential structures has been chosen. Does it mean that 
calculus, although very powerful is not necessary to describe physical 
phenomena? 
It might not be easy to find any answer to these questions.

 \ \ \ {\bf Acknowledgment:} I greatly enjoyed the hospitality 
 extended
to me during a stay at the Physics Department at the University of
Wisconsin-Madison, where the final version of the paper was discussed  
and written down.  
This work was supported in part by the grant 
{\bf KBN-PB 2253/2/91} by the {\bf 
University of Silesia} grant 
and by {\bf The II Joint M. 
Sk\l odowska-Curie USA-Poland Fund MEN-NSF-93-145}.

\newpage

\section*{References}
\newcounter{bban}
\begin{list}
{\ \arabic{bban} .\ }{\usecounter{bban}\setlength{\rightmargin}
{\leftmargin}}

\item M. Freedman, J. Diff. Geom. {\bf 17}, 357 (1982).
\item S. K. Donaldson, J. Diff. Geom. {\bf 18}, 279 (1983).
\item R. E. Gompf, J. Diff. Geom. {\bf 18}, 317 (1983).
\item S. DeMichelis and M. Freedman, J. Diff. Geom. {\bf 35}, 
219 (1992).
\item R. E. Gompf, J. Diff. Geom. {\bf 37}, 199 (1993).
\item C. H. Brans and D. Randall, Gen. Rel. Grav. {\bf 25}, 205 
(1993).
\item C. H. Brans, Class. Quantum Grav. {\bf 11}, 1785 (1994).
\item R. Wald, General Relativity, Chicago Univ. Press (1984).
\item H. Donnelly, Bull. London Math. Soc. {\bf 7}, 147 (1975).
\item M. Kreck and S. Stolz, Ann. Math. {\bf 127}, 373 (1988).
\item S. Stolz, Inv. Math. {\bf 94}, 147 (1988). 
\item R. Ma\'nka and J. S\l adkowski, Phys. Lett. {\bf B224}, 97 (1989).
\item R. Ma\'nka and J. S\l adkowski, Acta Phys. Pol. {\bf B21}, 
509 (1990).
\item A. Connes, in {\it The interface of mathematics and physics} 
(Claredon, Oxford, 1990) eds . D. Quillen, G. Segal and S. Tsou.
\item A. Connes and J. Lott, Nucl. Phys. {\bf B} Proc. Suppl.
{\bf 18B}, 29 (1990).
\item J. S\l adkowski, Acta Phys. Pol. {\bf B24}, 1255 (1994).
\item J. S\l adkowski, Int. J. Theor. Phys. {\bf 33}, 2381 (1994).
\item J. S\l adkowski, Madison Univ. preprint, MAD-PH-94-854 (1994).
\end{list}

\end{document}